\documentclass[useAMS,usenatbib,letterpaper]{mn2e}
\usepackage{amsmath,amssymb}
\usepackage{graphicx}
\usepackage{natbib}

\def\Omm{{\Omega_m}}
\def\Ommz{{\Omega_m^{\,z}}}

\def\Omk{{\Omega_k}}
\def\Oml{{\Omega_{\Lambda}}}

\def\aap{A\&A}
\def\apj{ApJ}
\def\aapr{A\&A Rev.}
\def\apjl{ApJ}
\def\mnras{MNRAS}

\def\aj{AJ}

\def\nat{Nat}

\def\apss{Ap\&SS}

\newcommand{\beq}{
\begin{equation}
}
\newcommand{\eeq}{
\end{equation}
}
\newcommand{\kms}{\,{\rm km\,s^{-1}}}
\newcommand{\msun}{\,{\rm M_\odot}}

\def\simlt{\mathrel{\rlap{\lower 3pt\hbox{$\sim$}}\raise 2.0pt\hbox{$<$}}}
\def\simgt{\mathrel{\rlap{\lower 3pt\hbox{$\sim$}} \raise 2.0pt\hbox{$>$}}}

\def\gsim{ \lower .75ex \hbox{$\sim$} \llap{\raise .27ex \hbox{$>$}} }
\def\lsim{ \lower .75ex\hbox{$\sim$} \llap{\raise .27ex \hbox{$<$}} }

\def\beq{\begin{equation}}
\def\eeq{\end{equation}}
\def\msun{$M_\odot$}

\def\Omm{{\Omega_m}}
\def\Ommz{{\Omega_m^{\,z}}}

\def\Omk{{\Omega_k}}
\def\Oml{{\Omega_{\Lambda}}}

\title[High redshift massive black holes]
{Assessing the redshift evolution of massive black holes and their hosts}  
\author[Volonteri \& Stark]
{M. Volonteri$^1$ \& D.P. Stark $^2$
\\
\\
$^1$Astronomy Department, University of Michigan, Ann Arbor, MI 48109 \\
$^2$Kavli Institute of Cosmology and Institute of Astronomy, 
Madingley Road, Cambridge CB3 0HA, UK\\}

\begin{document}  

\maketitle

\begin{abstract}

Motivated by recent observational results that focus on high redshift
black holes,  we explore the effect of scatter and observational
biases on the ability to recover the intrinsic properties of the black
hole population at high redshift.  We find that scatter and selection
biases can hide the intrinsic correlations between black holes and
their hosts, with `observable' subsamples of the whole population
suggesting, on average, positive evolution  even when the underlying
population is characterized by no- or negative evolution.  We create
theoretical mass functions of black holes convolving the mass function
of dark matter halos with standard relationships linking black holes with
their hosts.  Under these assumptions, we find that the local $M_{\rm BH}-\sigma$ correlation
is unable to fit the $z=6$ black hole mass function proposed by
Willott et al. (2010), overestimating the number density of all but
the most massive black holes. Positive evolution or including scatter
in the $M_{\rm BH}-\sigma$ correlation makes the discrepancy worse, as
it further increases the number density of observable black holes.  We
notice that if the $M_{\rm BH}-\sigma$ correlation at $z=6$ is steeper
than today, then the mass function becomes shallower. This helps reproducing the
mass function of $z=6$ black holes proposed by Willott et
al. (2010). Alternatively, it is possible that very few halos  (of
order 1/1000) host an active massive black hole at $z=6$, or that most
AGN are obscured, hindering their detection in optical surveys. 
Current measurements of the high redshift black hole
mass function might be underestimating the density of low mass black
holes if the active fraction or luminosity are a function of host or
black hole mass.  Finally, we discuss physical scenarios that can possibly lead
to a steeper  $M_{\rm BH}-\sigma$ relation at high redshift.
\end{abstract}

\section{Introduction}

The constraints on black hole masses at the highest redshifts
currently probed, $z\simeq6$, are few, and seem to provide conflicting
results.  (i) There seems to be little or no correlation between black
hole mass and velocity dispersion, $\sigma$ \citep{Wang2010} in the
brightest radio-selected quasars, (ii) typically black holes are
`over-massive' at fixed galaxy mass/velocity dispersion compared to
their $z=0$ counterparts \citep[e.g., Walter et al. 2004; at lower
  redshift see also][]{Mclure2004,Shields2006,Peng2006b, Decarli2010,
  Merloni2010, Woo2008}, but (iii) analysis of the black hole
mass/luminosity function  and clustering suggests that either many
massive galaxies do not have black holes, or these black holes are
less massive than expected \citep[W10 hereafter]{Willott2010}.  

As a result of point (ii), most authors propose that there is {\it
  positive} evolution in the $M_{\rm BH}-$galaxy relationships, and
quantify it as a change in {\it normalization}, in the sense that at
fixed galaxy properties (e.g. velocity dispersion, stellar mass),
black holes at high redshift are more massive than today.
 For
instance, 
Merloni et al. (2010) propose that $M_{\rm
  BH}-M_*$ evolves with redshift as $(1+z)^{0.68}$ while 
\cite{Decarli2010} suggest $(1+z)^{0.28}$.   Point (iii) above, 
however, is inconsistent with this suggestion unless only about 
$1/100$ of galaxies with stellar mass $\simeq 10^{10}-10^{11} $ \msun
~at $z=6$ host a black hole (W10). These galaxies are nonetheless
presumed to be the progenitors of today's massive ellipticals, which
typically host central massive black holes. 

When inferences on the population of massive black holes at the
highest redshift are made, we have to take into consideration two
important selection effects (see Lauer et al. 2007b). First, only the most massive black holes,
powering the most luminous quasars, can be picked up at such high redshifts
\citep{Shen2008, Vestergaard2008}. Second, as a result of the limited
survey area of current imaging campaigns, only black holes  that
reside in relatively common galaxies can be recovered. Taken together,
these biases imply  that the observable population of black holes at
high redshift will  span a narrow range of masses and host
properties (see also Adelberger et al. 2005, Fine et al. 2006). 

In this paper, we explore the impact of these observational biases 
on attempts to recover the intrinsic properties of the black hole population.
Our calculations are based on simple models grounded on empirical relations 
measured at much lower redshift, and therefore our results should be treated with caution. 
The aim of this paper is only to highlight  the effects of the different factors that can influence the 
measurement of the intrinsic properties of the black hole population at high redshift.

In section 2 we describe how we generate Monte Carlo realizations of
the $M_{\rm  BH}-\sigma$ relation at $z=6$  varying the slope and
normalization. We then select `observable' systems from these samples,
considering both `shallow' or `pencil beam' surveys, and test how
well we can recover  the parameters  of the $M_{\rm  BH}-\sigma$
relation from the `observable' systems.   In section 3 we discuss
theoretical mass functions of black holes derived from the mass
function of dark matter halos and various assumptions for the 
$M_{\rm BH}-\sigma$ relationship. Using these results, we test what assumptions
can reproduce the black hole mass function derived by W10.  We also discuss
(section 4) why obtaining constraints on the average accretion rates
and active fraction of black holes as a function of host mass is
crucial to our understanding of the high-redshift massive black hole
population. Finally, in section 5
we propose a simple theoretical framework that leads to selective
accretion onto black holes in a manner that reconciles the
observational results (i)-(ii) and (iii) above.


\section{Scatter and evolution of the $M_{\rm BH}-\sigma$ relation at high redshift}

We can  qualitatively show the effects of selection biases with a simple exercise. Let us 
assume an evolution of the $M_{\rm BH}-\sigma$ relationship of the form:
\beq
M_{\rm BH,\sigma}=10^8 \,{\rm M_\odot} \left( \frac{\sigma}{200 \kms} \right)^\alpha (1+z)^{\gamma},
\label{eq:MS_z}
\eeq
where $\alpha$ is a function of redshift.  Let us now also assume that
at fixed $\sigma$ the logarithmic scatter in black hole mass is $\Delta=$0.25-0.5 dex
($M_{\rm BH}=M_{\rm BH,\sigma}\times 10^{\Delta \delta}$, where $\delta$ 
is normally distributed, see, e.g., Gultekin et al. 2009, Merloni et al. 2010. 
The results are qualitatively unchanged for a uniform distribution in $\log\Delta$.)


%

We create a Monte Carlo simulation of the $M_{\rm BH}-\sigma$ relation
at $z=6$ assuming different values of $\alpha$ and
$\gamma$.   For this exercise we run a number of realizations $N(M_{h})
\propto 1/n(M_{h})$, where $n$ is the number density of halos of a given
mass ($M_h$) calculated through the Press \& Schechter formalism.   We then select only systems that are likely to be observed.   We consider a shallow survey and a pencil
beam survey.  A wide, shallow survey would preferentially select
systems with high luminosity, but has the advantage of a large area.
For instance the SDSS quasar catalogue selects sources with
luminosities larger than $M_i = -22.0$ ($\simeq 10^{45}$ erg s$^{-1}$)
over an area of 9380 deg$^2$, corresponding to a volume of almost 7
comoving Gpc$^3$ at z=6.  To simulate a shallow survey, we select
black holes with a sizeable mass, implying that large luminosities can
be achieved, $M_{\rm BH}>3\times 10^8$ \msun~ (see, e.g. Salviander et
al. 2007; Lauer et al. 2007b, Vestergaard et al. 2008,  Shen et
al. 2008, 2010 for a discussion of this bias), and hosted in halos with space
density $n>1$ Gpc$^{-3}$.  Pencil beam surveys can
probe fainter systems, but at the cost of a smaller area, e.g. the 2
Ms Chandra Deep Fields cover a combined volume of  $\simeq10^5$
comoving Mpc$^3$ at $z=6$ and reach flux limits of $\simeq 10^{-17}$
and $\simeq 10^{-16}$ erg cm$^{-2}$ s$^{-1}$ in the 0.5-2.0 and 2-8
keV bands, respectively (the flux limit corresponds to a luminosity
$\simeq10^{43}$ and $\simeq10^{44}$ erg s$^{-1}$at $z=6$). As an
example of a pencil beam survey, we select black holes with mass
$M_{\rm BH}>10^7$ \msun~ hosted in halos with  density $n>10^3$
Gpc$^{-3}$

To select sources that are observable in current surveys, we link the
velocity dispersion, $\sigma$, to the mass of the host dark matter
halo.   Empirical correlations have been found between  the central
stellar velocity dispersion and the asymptotic circular  velocity
($V_{\rm c}$) of galaxies (Ferrarese 2002;  Baes et al. 2003; Pizzella
et al. 2005).  Some of these  relationships (Ferrarese, Baes) mimic
closely the simple $\sigma=V_c/\sqrt[]{3}$ definition that one derives
assuming complete orbital isotropy. Indeed, it is difficult to imagine
that the ratio between  $\sigma$ and $V_c$ for massive, stable systems 
evolves strongly with redshift and that it can be much different from~
$\sqrt[]{3}$ or~$\sqrt[]{2}$ (see Binney \& Tremaine 2008).  
Since the asymptotic circular  velocity ($V_{\rm c}$) of galaxies is a measure
of the total mass of the dark matter halo of  the host galaxies, we
can derive relationships  between black hole and dark matter halo
mass, adopting, for instance,  Equation 1 with $\alpha=4$ and
$\gamma=0$: \beq M_{\rm h}=8.2\times10^{13} M_\odot \left[\frac{M_{\rm
      BH}}{10^9 M_\odot} \right]^{0.75} \left[ \frac{\Omm}{\Ommz}
  \frac{\Delta_{\rm c}} {18\pi^2} \right]^{-1/2} (1+z)^ {-3/2} .
\label{eq:sqrt3}
\eeq
In the above relationship, $\Delta_{\rm c}$ is the over--density at virialization relative  
to the critical density.  For a WMAP5 cosmology we adopt here the  fitting formula 
 $\Delta_{\rm c}=18\pi^2+82 d-39 d^2$ (Bryan \& Norman 1998),  where $d\equiv \Ommz-1$ 
 is evaluated at the redshift of interest, so that $ \Ommz={\Omm (1+z)^3}/({\Omm (1+z)^3+\Oml+\Omk (1+z)^2})$. 
 Given the mass of a host halo, we  estimate the number density from the Press \& Schechter formalism (Sheth \& Tormen
1999). In this section we assume $\sigma=V_c/\sqrt[]{3}$, where $V_c$ is the virial
circular velocity of the host halo. The results of this experiment are not strongly dependent on this specific assumption; 
  in Section 3 below we discuss different scalings. Kormendy et al. (2011) question a correlation between 
black holes and dark matter halos (but see Volonteri et al. 2011 for an updated analysis).
We notice that in any case Kormendy's argument is not a concern here, as at large masses 
Kormendy et al (2011b) suggest that a cosmic conspiracy causes $\sigma$ and $V_c$ to 
correlate, thus making the link between $M$ and $V_c$ adequate. 
In the Monte Carlo simulation, at fixed halo mass ($M_h$, hence, $\sigma$), 
we derive $M_{\rm BH,\sigma}$ from the adopted $M_{\rm BH}-\sigma$ relation (i.e. depending on 
the choice of $\alpha$ and $\gamma$), and then we draw the black hole mass
from $M_{\rm BH}=M_{\rm BH,\sigma}\times 10^{\Delta \delta}$ with varying values of the
scatter $\Delta$.
  
\begin{figure}
\includegraphics[width= \columnwidth]{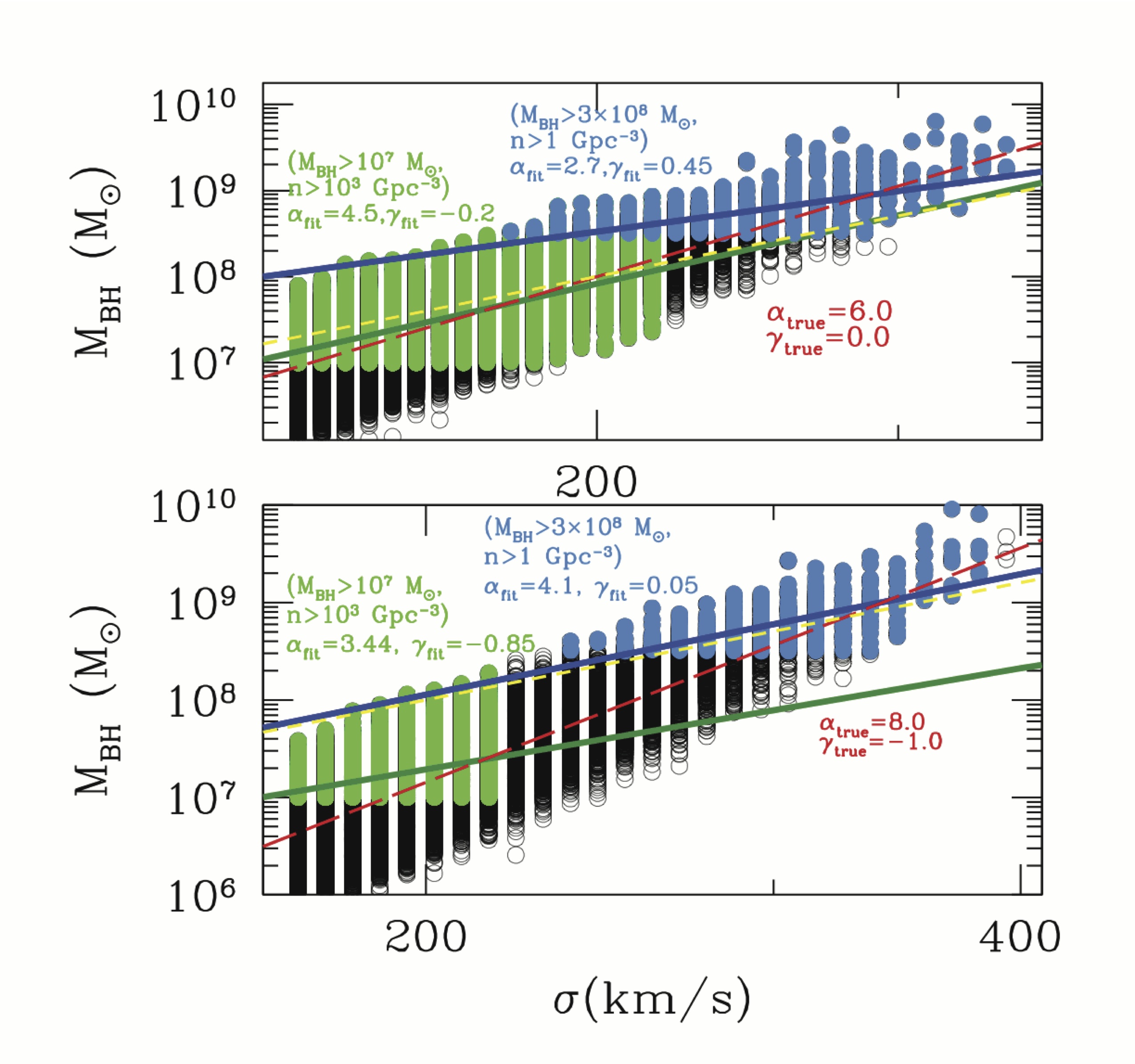}
\caption{Top panel: $M_{\rm BH}-\sigma$ relation at $z=6$, assuming $\alpha=6$, $\gamma=0$, and a scatter of 0.25 dex. Cyan dots: `observable' population in a shallow survey.  Blue line:  linear fit to this `observable'  population, yielding $\alpha=2.7$.
Green dots: `observable' population in a pencil-beam survey.  Dark green line:  linear fit to this `observable'  population, yielding $\alpha=4.5$.
Red line: fit to the whole population, yielding $\alpha=6$. 
Yellow dashed line: $M_{\rm BH}-\sigma$ at $z=0$ (Equation 1 with $\alpha=4$ and $\gamma=0$).  
Bottom panel: same for $\alpha=8$, $\gamma=-1$.}
\label{MS_z6}
\end{figure}

First, we test a no-evolution case, where we set $\alpha=4$ and
$\gamma=0$.  We fit, in log-log space, the $M_{\rm BH}-\sigma$ relation of black holes 
implied by the `observable'  population, considering both a shallow
and pencil beam survey. In the no evolution case, we find
$\alpha_{\rm fit} \simeq 1$ and $\gamma_{\rm fit} \simeq 0.7$ in the
`shallow' survey, with almost all `observable' black holes lying {\it
  above} the $\alpha=4$ and $\gamma=0$ line, suggesting `overmassive'
black holes, only because of the mass threshold that was imposed on
the sample. In the `pencil beam' survey we find $\alpha_{\rm fit}
\simeq 2.5$ and $\gamma_{\rm fit} \simeq -0.2$. In either case,
fitting only the `observable'  population yields a much shallower
slope than that characterizing the whole population.  

In Figure~\ref{MS_z6} we show a Monte Carlo simulation of the $M_{\rm
  BH}-\sigma$ relation at $z=6$ one would find assuming $\Delta=0.25$, $\alpha=6$ and
$\gamma=0$ (top panel), and $\alpha=8$ and $\gamma=-1$ (bottom panel).
In section 3, we will show that these particular choices of $\alpha$
and $\gamma$ are motivated by our attempt to fit the black hole mass
function of W10.  In the $\alpha=6$ and $\gamma=0$ Monte Carlo
simulation, we find that the best fit has $\alpha_{\rm fit} \simeq 2.7
\pm 0.2$ and $\gamma_{\rm fit} \simeq 0.45 \pm 0.04$ for the `shallow'
survey. The apparent normalization of the relationship therefore
increases by 0.35 dex (all the blue points lie above the yellow line
in the top panel of figure~\ref{MS_z6}). So while the underlying
population is characterized only by a change in slope (with respect to
the $z=0$ relationship), what would be recovered from the `observable'
population is a shallower slope and a positive evolution of the
normalization (in agreement with point (ii) in \S1).  We note,
additionally, that the smaller the range in $M_{BH}$ that is probed,
the more likely it is that the scatter $\Delta$ hides {\it any}
correlation, likely explaining the lack of correlation (point (i) in
\S1) found by Wang et al. (2010)\footnote{Wang et al. did not attempt
  any fit to the $M_{\rm BH}-\sigma$ relation. They note that they
  find significant scatter, extending to over 3 orders of magnitude,
  and that most of the quasar black hole masses lie above the local
  relationship. See also Shields et al. (2006) for quasars at
  $z=3$.}.  If we increase the level of scatter ($\Delta$) the slope of the 
  relationships recovered from the Monte Carlo sample becomes progressively shallower. 

We can repeat the same exercise for, e.g., $\alpha=8$ and $\gamma=-1$,
and although the underlying population has a much steeper slope and a
{\it negative} evolution of the normalization of the $M_{\rm
  BH}-\sigma$ relation with redshift, the `observable' population in
the  shallow survey would nevertheless display no evolution at all
(blue vs yellow lines in Fig.~\ref{MS_z6}). 

Summarizing we find that, (1) selection effects can severely  alter
the mapping between black mass and host galaxy velocity dispersion,
leading  to observed black hole populations that are more massive than
the true distribution.  (2) Scatter and selection effects can mask
correlations between black mass and  host galaxy properties, leading
to observed $M_{\rm BH}-\sigma$ relations that are shallower than the
true relation.  Although the quantitative results must be taken with caution,
the existence of biases towards measuring a positive evolution 
in the black hole-host correlations induced by selection and scatter 
is generically a robust result (e.g.,  Shields et al. 2006, Salviander et al. 2007; Lauer et al. 2007b).

\section{Impact of evolution of $M_{\rm BH}-\sigma$ relation and scatter on the black hole mass function}
We now turn to the mass function of black holes, and how its shape and
normalization are affected by the evolution of $M_{\rm BH}-\sigma$
relation and its scatter.  We create theoretical mass functions based
on Equation 1 coupled with the Press \& Schechter formalism, exploring
how different values of $\alpha$ and $\gamma$ influence its functional
form. As discussed in section 2, one can derive relationships between
black hole and dark matter halo mass given a relationship between
black hole mass and velocity dispersion (Equation 1),  a relationship
between velocity dispersion ($\sigma$) and asymptotic circular
velocity (virial velocity, $V_c$), and the virial theorem.  For
instance, assuming Equation 1 with $\alpha=4$ and $\gamma=0$, and
$\sigma=V_c/\sqrt[]{3}$  one derives Eq.~\ref{eq:sqrt3}, while if we
assume the relationship proposed by Pizzella et al. (2005) between
$\sigma$ and $V_c$: \beq M_{\rm h}=4.1\times10^{13} M_\odot
\left[\frac{M_{\rm BH}}{10^9 M_\odot} \right]^{0.56} \left[
  \frac{\Omm}{\Ommz} \frac{\Delta_c} {18\pi^2} \right]^{-1/2}  (1+z)
^{-3/2}.
\label{eq:pizzella}
\eeq

To consider the range of possible black hole mass functions, we adopt
the two mappings between black hole mass and halo  mass provided by
Equations~\ref{eq:sqrt3} and \ref{eq:pizzella}.   We first consider
the resulting black hole mass function when we adopt the local $M_{\rm
  BH}-\sigma$  relation ($\alpha=4$ and $\gamma=0$), and we will then
investigate how the mass function changes if we vary $\alpha$ and
$\gamma$.  In particular, we will focus on $\alpha=6$ and $\gamma=0$,
and $\alpha=8$ and $\gamma=-1$, because, as shown below, a steeper
$M_{\rm BH}-\sigma$  relation yields better agreement between
theoretical black hole mass functions and W10.

We can estimate the mass function of black holes by convolving
equations~\ref{eq:sqrt3}, \ref{eq:pizzella} (and their possible
redshift evolution)  with the mass density of dark matter halos with
mass $M_{\rm h}$ derived from  the Press \& Schechter formalism (Sheth
\& Tormen 1999): \beq \frac{dN}{d \log M_{\rm BH}}=\frac{dN}{d \log
  M_{\rm h}} \frac{d\log M_{\rm h}}{d\log M_{\rm BH}}.  \eeq

\begin{figure}
\includegraphics[width= \columnwidth]{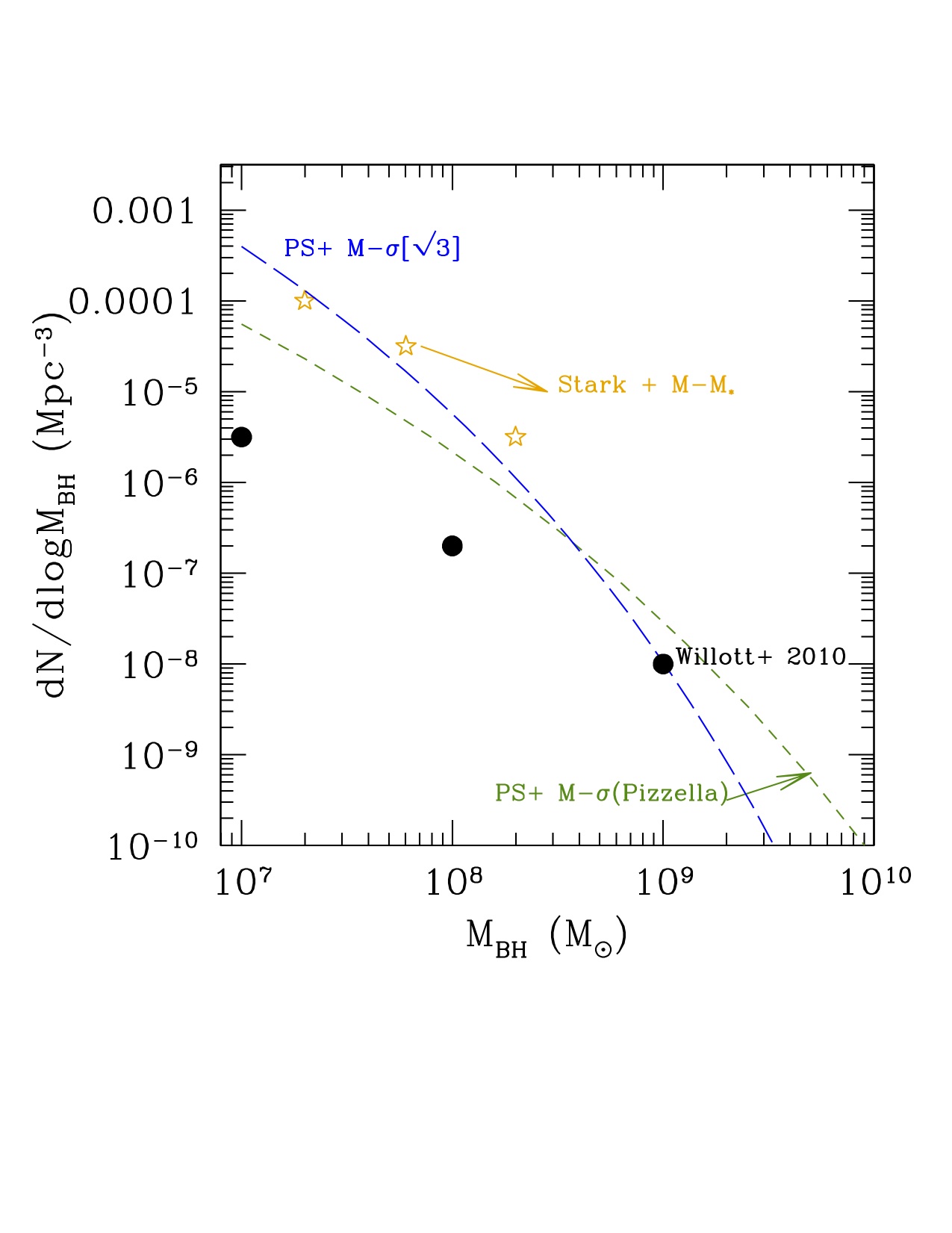}
\caption{Mass function of black holes. Black dots: Willott et al
  2010. Orange stars:  $M_{\rm BH}-M_*$ + Stark et al. 2009 (see
  Willott et al. 2010 for details).  Blue long dashed curve:  Press \&
  Schechter + equation~\ref{eq:sqrt3} ($\alpha=4$).  Dark green short
  dashed curve:  Press \& Schechter + equation~\ref{eq:pizzella}
  ($\alpha=4$).  }
\label{testz6_PS_w2}
\end{figure}

We assume for the time being that black holes exist in all
galaxies. The effect of  dropping this assumption is discussed in
detail in Section 4.  In figure~\ref{testz6_PS_w2} we compare the mass
function derived using this  technique to the mass function proposed
by W10, based on the luminosity function  of quasars selected by the
Canada-France High-z Quasar Survey, assuming a duty  cycle
(corresponding to the fraction of black holes that are active, we will
refer to this quantity as the active fraction, AF, below) of 0.75 and
assuming  a lognormal distribution of Eddington fractions, $f_{\rm
  Edd}$,  centered at  0.6 with standard deviation of 0.30 dex
\citep[see also][]{Shankar2010a}.   W10 further assume the same fraction of obscured AGN  
as observed at lower redshift ($z=0-2$, Ueda et al. 2003), and correct for  Compton thick AGN 
following Shankar et al. (2009).  Note that the evolution of the fraction of obscured or Compton thick AGN (currently not well constrained 
at high redshift, but see Treister et al. 2011) can strongly influence the results by hiding part of 
the black hole population (see section 4).

If the $M_{\rm BH}-\sigma$ relation evolved with redshift as proposed
by \cite{Woo2008}, $\gamma=3.1$, the number density of black holes in
the mass range $10^7-10^9$ M$_\odot$ would be $\simeq 0.5$ and
$10^{-4}$ comoving Mpc$^{-3}$ respectively (the curve corresponding to
this {\it very} strong evolution is not shown in the figure). We note,
however, that the sample analyzed by \cite{Woo2008} is at $z\approx
0.4$, and there is no guarantee that such evolution holds at higher
redshift.  

In all cases the analytical models greatly over-estimate the mass function at masses $M_{\rm BH}<10^9$ \msun, and
possibly at all masses -- when we add the suggested {\it positive}
redshift evolution of the $M_{\rm BH}-$galaxy relationships.

\begin{figure}
\includegraphics[width= \columnwidth]{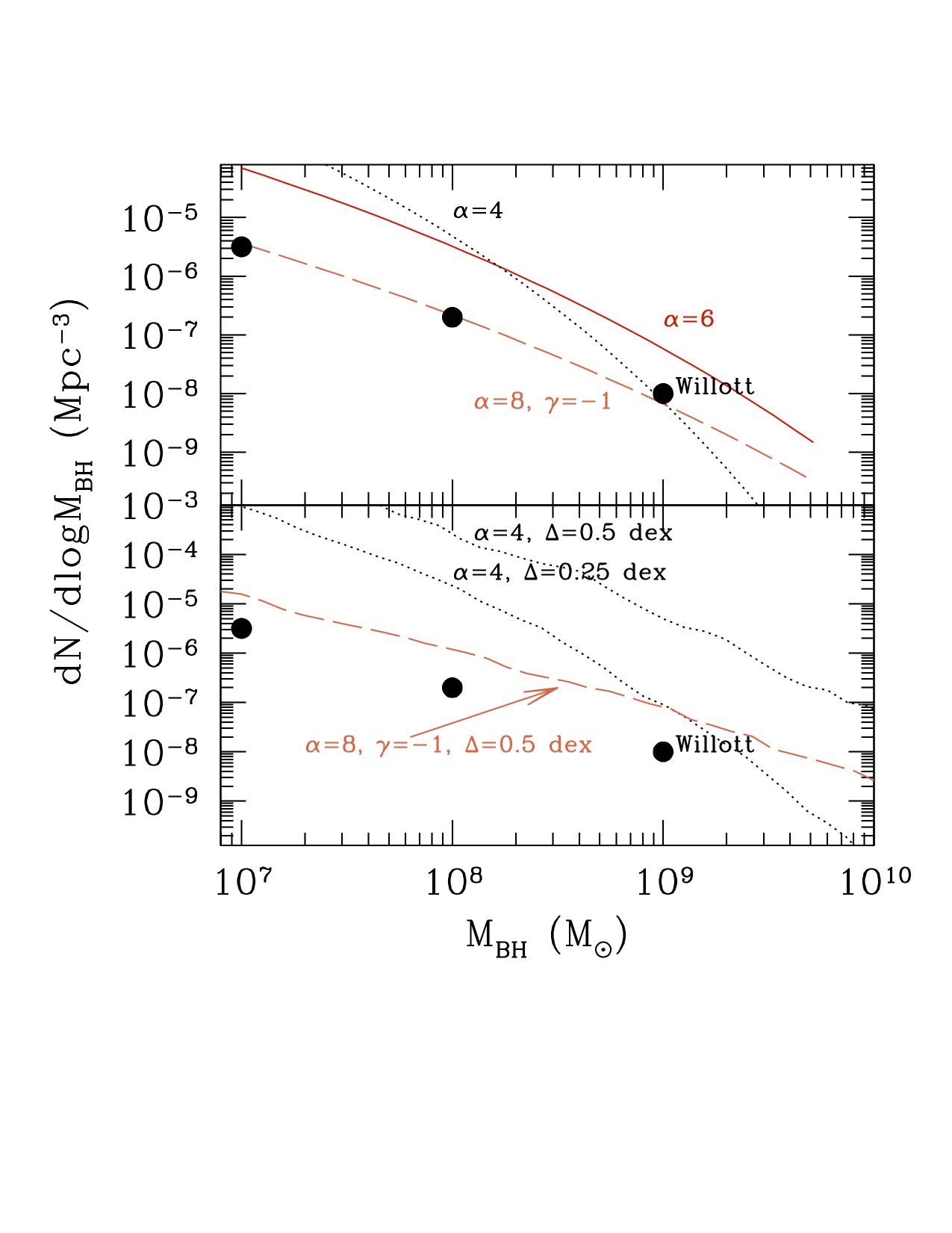}
\caption{Mass function of black holes. Top panel: we vary the slope
  ($\alpha$) and normalization ($\gamma$) of the $M_{\rm BH}-\sigma$
  relation, and assume no scatter in the relationship
  ($\Delta=0$). Bottom panel: we vary the slope of the $M_{\rm
    BH}-\sigma$ relation, and include scatter in the relationship
  ($\Delta=0.25$ dex or $\Delta=0.5$ dex).  Black dots: Willott et al
  2010.  All curves assume Press \& Schechter +
  equation~\ref{eq:sqrt3}, with varying $\alpha$ as labelled in the
  Figure.}
\label{testz6_PS_w1}
\end{figure}

In figure~\ref{testz6_PS_w1} we show instead the mass function we find
when we assume different  $\alpha$ and $\gamma$ values, with and
without scattering. We include scattering, at the level of
$\Delta=0.5$,  by performing a Monte Carlo simulation, where for each
black hole mass we create  500 realizations of the host mass.  The W10
black hole mass function can be reproduced by a simple model that has
$\alpha=8$ and $\gamma=-1$, if no or little scatter in the black hole
properties with galaxy mass is present. We see that as $\alpha$
increases the mass function becomes shallower.  At fixed black hole mass, 
above the `hinge' of Equation 1 (200 $\kms$) black holes will be found in comparatively 
less massive galaxies, that have a higher density.
On the other hand, below the `hinge', the host of a black hole of a given mass 
would be more massive than in in the $\alpha=4$ case, hence with a 
lower space density. This effect makes the mass function shallower. 
Any decrease in $\gamma$ tends to shift the black hole
mass function to lower number densities at all masses.  

However, a significant amount of scatter increases the number density
of observable black holes, as shown in the bottom panel of
Figure~\ref{testz6_PS_w1}.  This effect has been discussed extensively
by \cite{Lauer2007b}, and we refer the reader to this paper for an
exhaustive demonstration of its consequences.   \cite{Lauer2007b}
start from the luminosity function of galaxies, rather than the mass
function of dark matter halos, and the fact that the most luminous galaxies are
in the exponential part of the luminosity function, implies that the scattering of very
high-mass black holes ($M_{\rm BH}\simeq 10^9$ \msun) in lower mass
galaxies has a stronger effect than the scattering of low mass black
holes in larger galaxies.  A similar
conclusion applies to the mass function of dark matter
halos. Additionally, since the halo mass function becomes exponential
at lower masses at high redshift, the effect of scatter on the shape
of the black hole mass function becomes noticeable already at $M_{\rm
  BH}\simeq 10^7$ \msun.  Including scatter, the simple model with
$\alpha=8$ and $\gamma=-1$ is now a much poorer fit to W10 mass
function, but it still reproduces their slope very well. 

Summarizing, we find that the local $M_{\rm BH}-\sigma$  relation ($\alpha=4$ and $\gamma=0$) 
is unable to reproduce W10 results, even more so when a level of scatter compatible with observational results
($\Delta=0.25-0.5$) is included.  A steeper $M_{\rm BH}-\sigma$
relation, with possibly a negative evolution (e.g., $\alpha=8$ and
$\gamma=-1$) provides a better fit, although high levels of scatter
require an even more dramatic steepening of the slope in order to
match the mass function proposed by W10. 
 While the direct comparison with W10 strongly depends on the limitations of our empirical model,
the relationship between increased scatter in the $M_{\rm BH}-\sigma$ and increased number density 
of black holes is a robust result that directly follows from the analysis presented in  \cite{Lauer2007b}.

 \section{Occupation fraction of quiescent and active black holes}
In section 3 we demonstrated how we derive the theoretical mass function of
black holes from the mass function of their host halos and the
relation between black hole and halo masses
\citep[e.g.,][]{Haiman1998,Wyithe2002}.  However, when we convolve
equations~\ref{eq:sqrt3} and \ref{eq:pizzella}  with
the mass density of dark matter halos to derive a black hole mass
function, we have to make a conjecture about the fraction of halos 
of a given mass which host a black hole, the occupation fraction (OF):

\beq \frac{dN}{d \log M_{\rm BH}}={\rm OF}(M_{\rm h},z)\frac{dN}{d
  \log M_{\rm h}} \frac{d\log M_{\rm h}}{d\log M_{\rm BH}}. \eeq 

\begin{figure}
\includegraphics[width= \columnwidth]{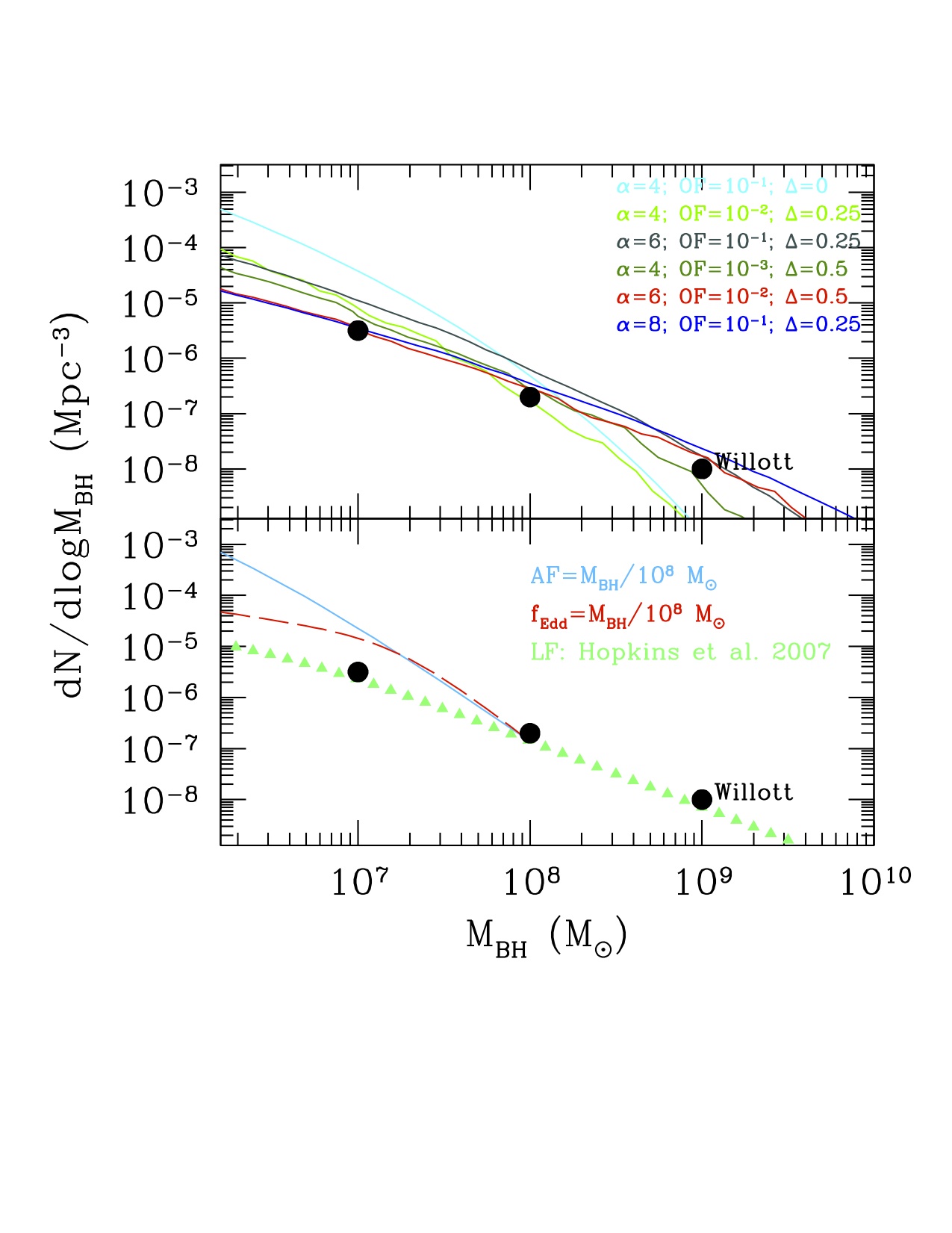}
\caption{Top: theoretical mass function of black holes derived from the mass function of dark matter halos (Eq.~5). Black dots: Willott et al 2010. Other curves as marked in the figure (from top to bottom). Willott et al. results can be reproduces for a range of possible assumptions on the relation between holes and halos.  Bottom: empirical mass function of black holes derived from the luminosity function of quasars (Eq.~7 and Eq.~8, adopting the bolometric luminosity function of Hopkins et al. 2007). Triangles: $f_{\rm Edd}=1$ and $AF=1$.  The blue solid and red dashed curves show how mass-dependent luminosities or active fractions can modify the shape of the mass function one would infer. }
\label{testz6_PS_w3}
\end{figure}

In the top panel of Fig.~\ref{testz6_PS_w3} we show black hole mass
functions resulting from different choices of the OF. It is clear that
a decreasing OF can compensate for an increased scatter in shaping the
black hole mass function.  As a result of this degeneracy, we can
reproduce the W10  mass function for a range of values for  the OF and
scatter. Even adopting $\alpha=4$ (slope of $M_{\rm BH}-\sigma$
relation   at the present day) with a sensible scatter can fit the
data, at the cost,  however, of making the presence of a black hole
(regardless of its  shining as a quasar) a very rare instance.  
We note that it is conceivable that the OF is not constant over all
host masses, and a non trivially constant OF is expected particularly
at high-redshift, close to the epoch of galaxy and black hole
formation \citep{Menou2001}. At face value, the W10
data can be reproduced by OF$=M_{\rm h}/5\times10^{13}$ \msun for
$\alpha=4$, $\gamma=0$ and $\Delta=0.25$, or OF$=(M_{\rm
  h}/10^{13}$\msun$)^{1.25}$ for $\alpha=4$, $\gamma=0$ and
$\Delta=0$. Such occupation fractions are several orders of magnitude
lowers than predicted by models of formation and cosmic evolution of
black holes \citep[e.g.,][]{Volonteri2010}, and we therefore still
prefer solutions with steeper slopes. We will explore
self-consistently OF and its relationship with the establishment of
the $M_{\rm BH}-\sigma$ relation as a function of black hole formation
and growth physics in a future paper. 

Throughout this paper we have compared our theoretical mass function
of black holes to constraints derived indirectly from the luminosity
function of quasars in W10, rather than from direct black hole mass
measurements (that are rather unfeasible at $z=6$). Empirically, one
can derive the mass function of black holes from the luminosity
function of quasars and a relation between black hole mass and quasar
luminosity \citep[e.g.,][]{Shankar2010a,Shankar2010b,Willott2010}:
 
 \begin{equation}
\frac{dN}{d \log M_{\rm BH}}=\frac{dN}{d \log L} \frac{d \log L}{d
  \log M_{\rm BH}}.
\end{equation}

For instance, we can estimate the mass function of black holes  from
the bolometric luminosity of radio-quiet quasars \citep{Hopkins2007}
assuming (1) that all black holes are active, (2) that all black holes
radiate at the same Eddington fraction, $f_{\rm Edd}$ (based on
various observational results we expect high redshift quasars to
radiate close to the Eddington limit, see W10 and references
therein). The mass of a black hole powering a quasar with luminosity
$L$ is then:

\beq \frac{M_{\rm BH}}{10^9 M_\odot}=3\times 10^{-14}\frac{1}{f_{\rm
    Edd}}\frac{L}{L_\odot},
\label{eq:fedd}
\eeq

and one can trivially turn the luminosity function into a mass
function. As discussed by W10, their mass function is derived assuming 
similar values of the Eddington ratio and the active fraction 
using a more accurate technique \citep[see][for
  details]{Shankar2010a}. Our simple approach provides results
consistent with W10 if we assume a constant $f_{\rm Edd}=1$.   
  
When we deconvolve the luminosity function of quasars to derive the
black hole mass function we have to assume an active fraction (AF):

\begin{equation}
\frac{dN}{d \log M_{\rm BH}}={\rm AF}(M_{\rm BH},z) \frac{dN}{d \log
  L} \frac{d \log L}{d \log M_{\rm BH}},
\end{equation}

where we  indicate that both the active fraction and the Eddington
ratio can be functions of the black hole (and host) properties, and of
cosmic time. 

The intrinsic shape of the mass function changes as a function of AF
and $f_{\rm Edd}$, and, in particular, any departure from the
assumptions  $f_{\rm Edd}=1$ and AF$=1$ (that are upper limits to both
quantities) will drive the mass function of black holes `up', that is,
they will increase the number of black holes at a given mass. We have
therefore to bear in mind that the semi-empirical mass function
derived by W10 might be underestimating the mass function. The lower
panel of  Fig.~\ref{testz6_PS_w3} shows how the mass functions one
derives from a luminosity function are modified by an AF or $f_{\rm
  Edd}$ that depend  on the BH mass. For instance, we can trivially
assume that $f_{\rm Edd}=M_{BH}/ 10^8$ \msun~ for $M_{BH}< 10^8$
\msun, and $f_{\rm Edd}=1$ otherwise. Then $M_{\rm BH}=10^8M_\odot
(L/3\times10^{12}L_\odot)^{0.5}$, using equation~\ref{eq:fedd} in the
last expression. In the same figure we also show the effect of a mass
dependent AF, where we adopt the simple expression AF$=M_{BH}/ 10^8$
\msun~ for $M_{BH}< 10^8$ \msun, and AF$=1$ otherwise. These specific
forms of the mass dependence of AF and $f_{\rm Edd}$ are motivated by
the expectation that the most massive black holes at the earliest
cosmic times are all actively and almost constantly accreting
\citep{Haiman2004,Shapiro2005,Volonteri2005,Volonteri2006}.
 Such functional forms are here only used to prove that non constant accretion rates 
modify the expectations in  terms of mass function and active fraction, but the 
expressions we adopt should be considered representative of any class of accretion 
rates and active fractions that are not constant, rather than actual predictions.

\begin{figure}
\includegraphics[width= \columnwidth]{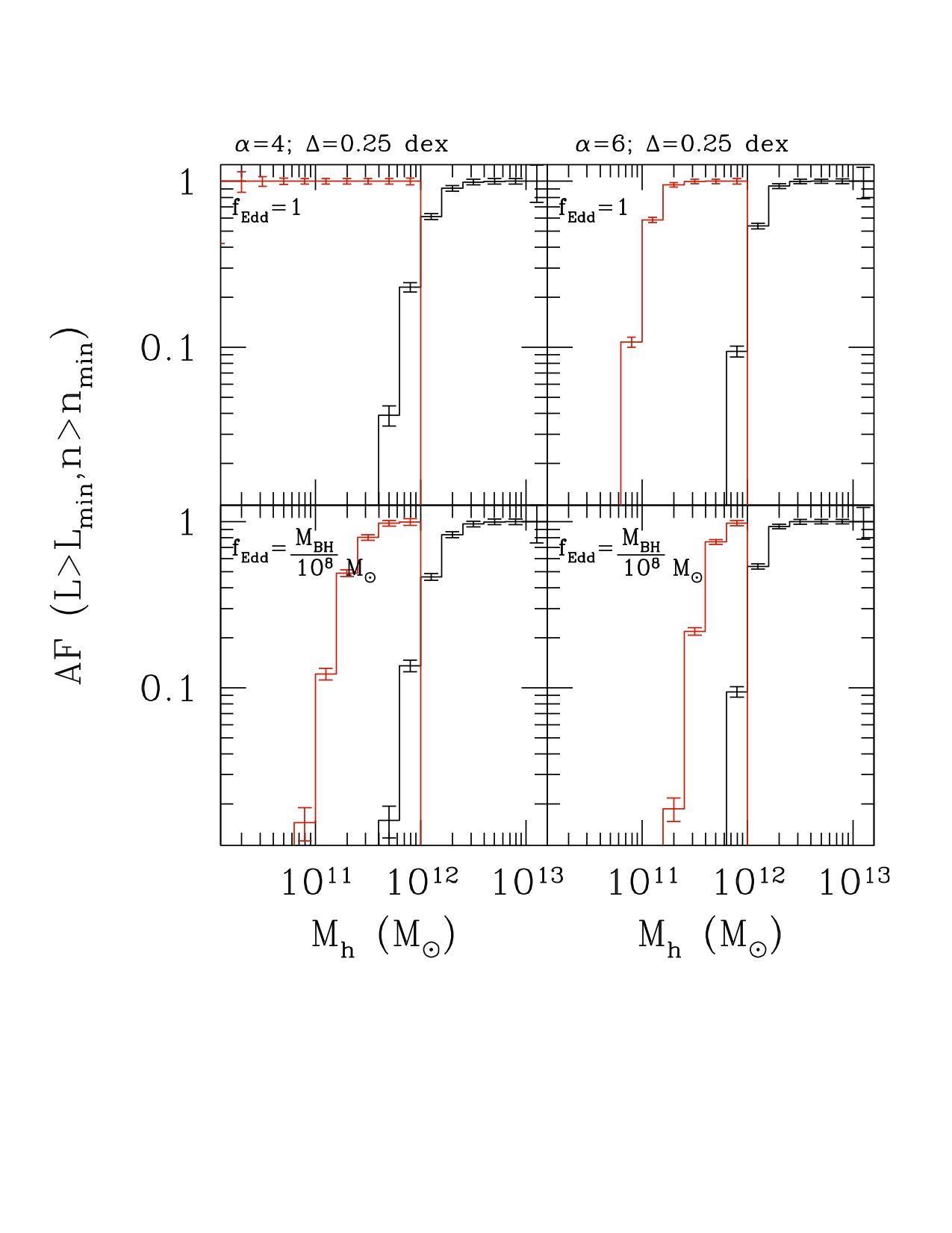}
\caption{Fraction of active black holes than can be detected in a survey with given luminosity and volume limit. Red histograms (leftmost side of each panel): black holes with $L_{\rm min}=10^{44}$ erg/s and $n_{\rm min}=10^{-5}$ Mpc$^{-3}$ (example of a pencil beam survey). Black histograms (rightmost side of each panel): black holes with $L_{\rm min}=10^{46}$ erg/s and $n_{\rm min}=10^{-9}$ Mpc$^{-3}$ (example of a shallow survey). Left panels: $\alpha=4$; $\Delta=$0.25 dex. Right panels: $\alpha=6$; $\Delta=$0.25 dex. Top panels: $f_{\rm Edd}=1$. Bottom panels: $f_{\rm Edd}=M_{BH}/ 10^8$ \msun.}
\label{active_alpha}
\end{figure}

If the Eddington ratio and/or AF are a function of halo or black hole
mass, then what one derives from flux-limited surveys will be
dependent on a combination of various
properties. Figure~\ref{active_alpha} shows simple examples. We build
a sample of black holes and hosts by performing a Monte Carlo sampling
as described in Section 3 for two $M_{\rm BH}-\sigma$ relations
($\alpha=4$ and $\alpha=6$, both with $\gamma=0$) each with a scatter
of 0.25 dex. We assign to each black hole a luminosity by assuming
either a constant $f_{\rm Edd}=1$, or that $f_{\rm Edd}=M_{BH}/
10^8$~\msun~ or $M_{BH}< 10^9$ ~\msun~ and $f_{\rm Edd}=$1 for
$M_{BH}> 10^8$ ~\msun (we note that the assumption that $f_{\rm Edd}$
scales with the halo mass, rather than the black hole mass, yields
very similar results). Even for a constant Eddington ratio, scatter
in the  $M_{\rm BH}-\sigma$ relation implies that at fixed halo mass
the black hole mass and hence its luminosity  is not univocally
determined. The `observed' active population of black holes  will
therefore be different from the `intrinsic' active fraction, which in
this exercise was set to unity. A comparison between the left and
right panels underscores how the $M_{\rm BH}-\sigma$ relation itself
shapes the fraction of active black holes which are detected in 
optical imaging surveys.

We notice that obscuration plays a role similar to the occupation or
active fraction. If, say, all black holes are active, but a large
fraction are Compton thick, then a large population of obscured
quasars would be unaccounted for in optical quasar surveys. There is
indeed evidence for a large fraction of high redshift quasars being
obscured \citep[e.g.,][]{lafranca2005,Treister2009,Treister2010}.



\section{Accretion efficiency and host mass}

In the previous section we discussed how the `observed' fraction of
active black holes, which goes into determining the `observed' mass
function depends on  the $M_{\rm BH}-\sigma$ relationship and on the
link between accretion rate and black hole-host masses. In this
section,  we explore the consequences and likelihood of a galaxy
mass-dependent black  hole accretion rate.  This hypothesis is
plausible, as the gas supply, especially at high redshift is likely
dependent on the environment and mass of the host. For instance cold
gas that flows rapidly to the center of galaxies from filaments around
haloes plays a major role in the buildup of massive galaxies at high
redshift \citep{Brooks2009,Governato2009}, with a transition expected
to occur when a galaxy has mass above $10^{11}$\msun, where gas is
shocked before it can reach the galaxy's disk. Cold gas flowing into
halos along large-scale structure filaments may however be dense
enough to penetrate the shock front and deliver cold gas to the
galaxy. Galaxies that form within a gas-rich filament will accrete gas
from this cold flow and grow substantially before the filament
dissipates. These galaxies embedded in filaments are expected to be
high peaks of the density field, hence among the most massive at early
times. 

Additionally, as discussed in section 3, instead of an overall
normalization evolution, the link between black holes and their hosts might
be better explained by an evolution in the slope of the $M_{\rm
  BH}-\sigma$ relationship. We are not claiming here that the 
evolution has the exact form that we use in this paper
(Equation~\ref{eq:MS_z}). We here discuss a possible physical scenario
that can lead to a steeper $M_{\rm BH}-\sigma$ relation.

In the following toy model we just explore what physical process could drive the establishment of a 
given  $M_{\rm BH}-\sigma$ relation at a given redshift ($z=6$ in this particular case). 
In other words, {\it if} the slope of the $M_{\rm BH}-\sigma$ relation at $z=6$ has a given $\alpha$, 
what can be the driver of such correlation? 

Let us assume that all black holes start with the same initial mass,
$M_0$, and let them grow until $z=6$ ($t_{H}=0.9$ Gyr). If at $z=6$
the slope of the $M_{\rm BH}-\sigma$ relationship is $\alpha$, and we
assume that on average black holes accrete at a fraction $\langle f_{\rm Edd} \rangle$ of the
Eddington rate,  then we can relate this average accretion rate of the
mass of the hole, $M_{\rm BH,6}$ and the velocity dispersion, $\sigma_6$, of the host at $z=6$, as follows: \beq
M_{\rm BH,6}=M_0 \exp\left({\langle f_{\rm Edd} \rangle\frac{t_H}{t_{\rm
      Edd}}\frac{1-\epsilon}{\epsilon}}\right),
\label{eq:BHdeltat}
\eeq

where $t_{\rm Edd}=M_{\rm BH} c^2/L_{\rm Edd}=\frac{\sigma_T \,c}{4\pi \,G\,m_p}= 0.45$ Gyr ($c$ is the speed of light, $\sigma_T$ is the Thomson cross section, $m_p$ is the proton mass), and the radiative efficiency, $\epsilon \simeq 0.1$. We can also express the relationship between black hole mass $M_{\rm BH,6}$ and $\sigma_6$ at $z=6$ as:

\beq
\sigma_6=200 \kms \left( \frac{M_{\rm BH,6}}{10^8 \,{\rm M_\odot}} \right)^{1/\alpha},
\eeq

so that the average accretion rate for a black hole that grows within a galaxy that has 
a velocity dispersion  $\sigma_6$ at $z=6$ results:

\beq
\langle f_{\rm Edd} \rangle=\frac{t_{\rm Edd}}{t_H} \frac{\epsilon}{1-\epsilon} \ln \left[ \left(\frac{10^8 \,{\rm M_\odot}}{M_0}\right)\left( \frac{\sigma_6}{200 \kms} \right)^\alpha \right].
\label{alphabh}
\eeq
Equations  9--11 are based on the initial conditions and on the properties at $z=6$ only. The accretion rate is in principle galaxy--mass dependent, and specifically dependent also on 
the mass growth of the host.  However, for the sake of simplicity it is here set to the average over the integration time.

\begin{figure}
\includegraphics[width= \columnwidth]{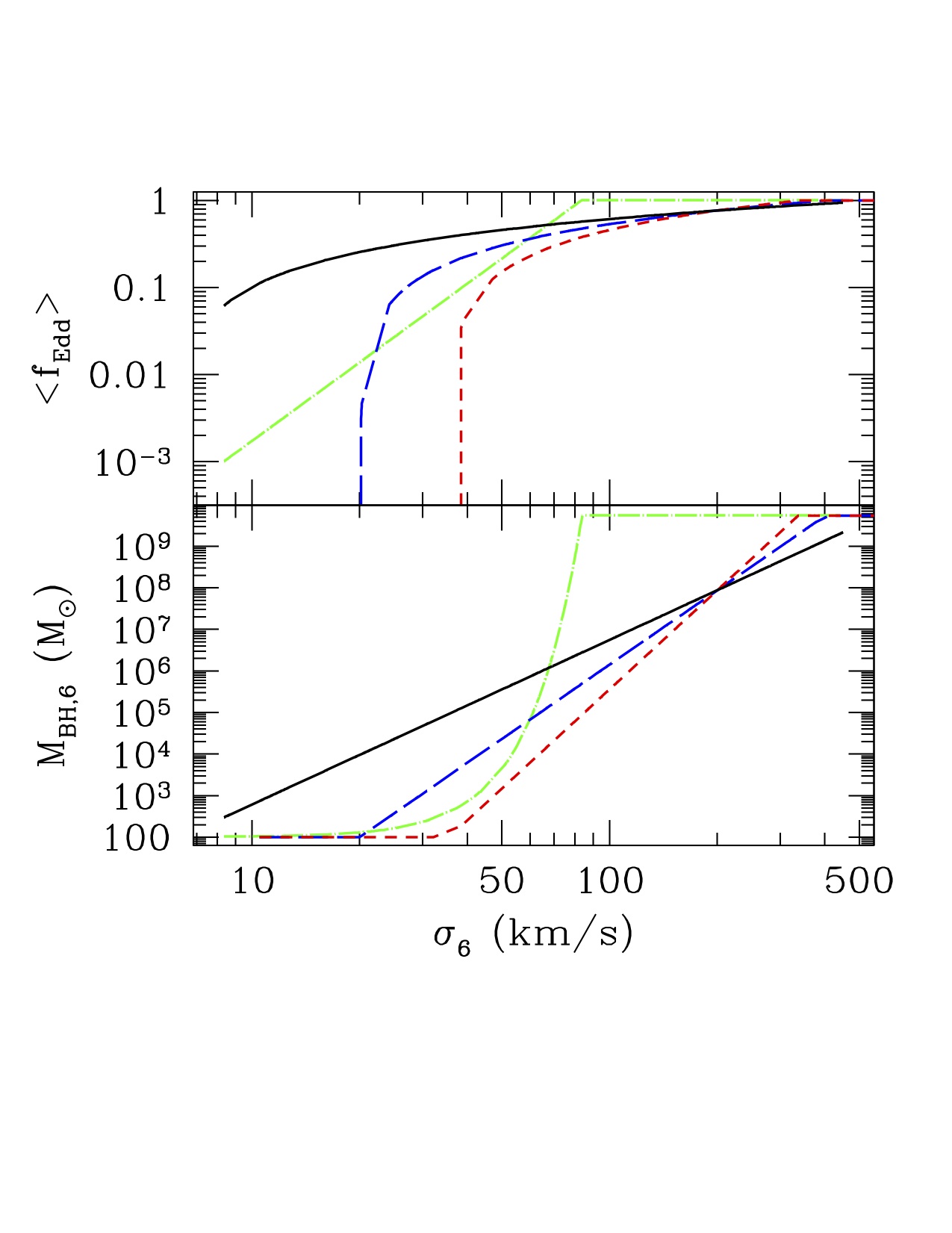}
\caption{Top: Eddington fraction as a function of halo velocity
  dispersion, derived from equation~\ref{alphabh} (black solid:
  $\alpha=4$; blue long dashed: $\alpha=6$; red short dashed:
  $\alpha=8$) and equation ~\ref{PP} (green dot-dashes). Bottom: black
  hole mass versus $\sigma$ at $z=6$ assuming $f_{\rm Edd}$ is a function of
  halo properties, and letting the holes accrete for $t_{H}=0.9$ Gyr. }
\label{f_Edd_fit}
\end{figure}

The average accretion rate of Eq.~11 is shown in Figure~\ref{f_Edd_fit} for $\alpha=4$,
$\alpha=6$ and $\alpha=8$. Figure~\ref{f_Edd_fit} implies that if only
black holes in hosts above a certain velocity dispersion, or mass, or
depth of the potential well, can accrete efficiently, it is only
natural to expect a different slope of the $M_{\rm BH}-\sigma$
relationship in dependence of the exact threshold. One possibility is
that black hole growth is indeed inefficient in low-mass galaxies at
early cosmic times, because of the fragile environment where  feedback
can be very destructive
\citep{Milos2009,Alvarez2009,Park2010,Johnson2010}. We note that all
these possible scalings of accretion rate with halo mass are
consistent with the independence of $f_{\rm Edd}$ on luminosity (and
black hole mass) found by W10, as all their black holes have masses
above $10^8\,{\rm M_\odot}$, where it is indeed expected that the
accretion rate can reach $f_{\rm Edd}\simeq1$ as one can infer from
Figure~\ref{f_Edd_fit}.

This exercise is not meant to suggest that the typical accretion 
rate has the exact value of Eq. 9, but that  {\it if} accretion is more efficient 
in more massive halos, then $\alpha$ increases, while if the accretion rate is mass 
independent, e.g. is constant in all hosts,  then $\alpha$ tends to lower values. 

 Equation~\ref{alphabh} simply demonstrates mathematically that 
in order to achieve a very steep $M_{\rm BH}-\sigma$ accretion in massive haloes has to be more efficient 
than in small haloes (`selective accretion'), and it should not be used to make predictions about the accretion/growth history of black holes.
To test this suggestion, dedicated simulations that can resolve the growth 
of black holes in cosmological simulations as a function of the host mass are 
required. However, simulations that explore the cosmic evolution of accretion efficiency,
taking into consideration feeding and feedback, as a function of host
mass at sufficient resolution are not currently available.

The experiment that is closest in spirit to what we propose was performed
by \cite{Pelupessy2007}\footnote{Indirectly, similar information can
  be extracted from \cite{Sijacki2009}, although all their information
  on the accretion rate is cast in terms if black hole masses, rather
  than host properties.}  who suggest that the more massive the host
halo, the higher the Eddington fraction.  A very simple fit from their
simulation results at $z=6$ (Figure 7 in Pelupessy et al. 2007) gives:

\beq f_{\rm Edd} \approx\frac{M_{\rm h}}{10^{11} \,{\rm
    M_\odot}}.
\label{PP}
\eeq

This equation is shown in Figure~\ref{f_Edd_fit} for different hosts,
where we assumed $\sigma=V_c/\sqrt[]{3}$, and that the Eddington limit
is capped at $f_{\rm Edd}=1$ (in the bottom panel the black hole mass
uses Equation~\ref{eq:BHdeltat}). This scaling leads to very steep
relationships between black holes and hosts, as shown in the bottom
panel of Figure~\ref{f_Edd_fit}, as the black hole mass depends 
exponentially on the halo mass (if we insert equation 12 into equation 9)
or on the time evolution of the halo mass (if we insert equation 12 into the 
expression for the accretion rate in Eddington units: 
$\dot M=f_{\rm Edd}(M/t_{\rm Edd})[(1-\epsilon)/\epsilon]=(M_{\rm h}(t)/10^{11})(M/t_{\rm Edd})[(1-\epsilon)/\epsilon]$. 
 To integrate this equation properly one needs to know the growth rate of the host dark matter halo mass as a 
function of time in a $\Lambda$CDM cosmology, $M_{\rm h}(t)$. Such exercise requires either merger trees that track the cosmic history
of dark matter halos, or an analytical fit to their growth histories, and it is beyond the scope of this paper). 
{\it Selective} accretion, modulated
by the host's potential and environment, is a possible key to
explaining a shallow high redshift black hole mass function without
requiring an unrealistically low occupation fraction.  

At late cosmic times we expect the interaction of black holes and galaxies to become 
more closely linked to baryonic processes (e.g., bulge formation) rather than being related
to the halo mass. For instance, secular effects might at late times decouple the 
properties of the central stellar-dominated region from the overall dark matter
halo. As another example, gas accretion through cold flows in filaments is expected to 
occur only at early times. We can, for instance, see a parallel between the black hole-halo relationship
and the baryon-halo relationship. It is expected that at very high redshift most halos
possess a baryon fraction of order the cosmic baryon fraction, while at later times the baryonic 
content evolves under the effect of baryonic physics. In the same way at late times we expect
black hole growth to be more closely related to the baryonic content of a galaxy, and hence less 
``selectively" linked to the host halo (cf. Volonteri et al. 2011).

%

\section{Summary and conclusions}

Recent observational results that focus on high-redshift black holes
provide seemingly conflicting results. In particular:
\begin{itemize}
\item[(i)] there seems to be little or no correlation with velocity
  dispersion, $\sigma$ in the brightest radio-selected quasars, 
\item[(ii)] typically black holes are `over-massive' at fixed galaxy
  mass/velocity dispersion compared to their $z=0$ counterparts, 
\item[(iii)] clustering and analysis of the mass/luminosity function
  suggest that either many massive galaxies do not have black holes,
  or these black holes are less massive than expected.
\end{itemize}

To try and understand these observational results, we explore the role of scatter and
observational biases in recovering the intrinsic properties of the
black hole population.  We generate Monte Carlo realizations of the
$M_{\rm  BH}-\sigma$ relation at $z=6$, varying the slope and
normalization, and select `observable' systems from these samples,
considering either `shallow' or `pencil beam' surveys. We test
how well we can recover the parameters  of the $M_{\rm  BH}-\sigma$
relation from the `observable' systems only.  We then create
theoretical mass functions of black holes from the mass function of
halos and $M_{\rm BH}-\sigma$ and test what assumptions can reproduce
the mass function derived by W10. 
 Our techniques are very simplified and we use empirical correlations that are not 
guaranteed to hold at all masses and reshifts.  Therefore one should not interpret our 
results as solutions to the three conflicting points mentioned above, but rather regard 
them as a way for understanding how different physical parameters may affect black hole 
related quantities and their measurements.

Our results can be summarized as follows:

\begin{itemize}
\item Scatter and bias selections can hide the intrinsic correlations
  between holes and hosts. When selecting within a small range in
  black hole and galaxy masses, at the high-mass end, the scatter
  washes out correlations (see point (i) above), and most of the
  selected systems will tend to lie above the underlying
  correlation. The correlations recovered from `observable'
  sub-samples of the whole population can therefore suggest positive
  evolution  even when the underlying population is characterized by
  no or negative evolution.
\item The slope and normalization of the local $M_{\rm BH}-\sigma$
  correlation are unable to produce a black hole mass function
  compatible with W10, as the theoretical mass function greatly
  overstimates the density of black holes with $M_{\rm BH}<10^8$. 
  The discrepancy can be minimized if very few halos (of order 1/1000) host a massive
  black hole or an AGN at $z=6$, or most AGN at these redshift are
  obscured.
  \msun.
\item If the $M_{\rm BH}-\sigma$ correlation were steeper at $z=6$, then 
at fixed black hole mass high mass black holes would reside in comparatively 
less massive galaxies than in the $\alpha=4$ case. Their number density is therefore 
increased. Viceversa, low mass black holes would be hosted in comparatively larger galaxies 
(compare red and yellow lines in Figure~1) with a lower space density. 
This effect helps reproducing the mass function of $z=6$ black holes proposed by W10. 
\item On the other hand, scatter in the $M_{\rm BH}-\sigma$, at the
  level of what is observed locally, exacerbates the discrepancy, as
  it increases the number density of black holes at $M_{\rm BH}>10^7$
  \msun. Any type of positive evolution of the $M_{\rm BH}-\sigma$
  exacerbates this discrepancy.
\item Analysis of AGN samples might be underestimating the black holes
  mass function at low masses if the active fraction or luminosity are
  a function of host or black hole mass.
\end{itemize}

 In the near future the synergy of {\it JWST} and {\it ALMA} can zoom
 in on quasars and their hosts respectively informing us of their
 relationship and how the $M_{\rm BH}-\sigma$ relation
 is established, or how the accretion properties depend on the black
 hole or halo mass.  In the near-IR, {\it JWST} will have the
 technical capabilities to detect quasars at $z\simgt 6$ down to a
 mass limit as low as $10^5-10^6$ \msun, owing to its large field of
 view and high sensitivity. At the expected sensitivity of {\it JWST},
 $\simeq 1$ nJy, almost $7\times 10^3$ deg$^{-2}$ sources at $z> 6$
 should be detected \citep{Salvaterra07}.  At the same time, the
 exquisite angular resolution and sensitivity of {\it ALMA} can be
 used in order to explore black hole growth up to high redshift even in
 galaxies with high obscuration and active star formation. To date the
 best studies of the hosts of $z\simeq 6$ quasar have been performed
 at cm-wavelength \citep{Walteretal2004,Wang2010}. The best studied
 case is J1148+5251 at z = 6.42.  The host has been detected in thermal 
dust, non-thermal radio
 continuum, and CO line emission
 \citep{Bertoldi2003,Carilli2004,Walteretal2004}.  {\it ALMA} will be
 able to detect the thermal emission from a source like J1148+5251 in
 a few seconds at sub-kpc resolution \citep{Carilli2008}. 
 On a similar time-frame Dark-Energy oriented
 survey will provide an enormous amount of quasar data as ancillary
 science (e.g. DES, LSST). Coupling the information we derive from
 these extremely large yet shallow surveys, with that derived from deep pencil 
 beam surveys will undoubtedly deepen our understanding of the growth of
 high-redshift black holes. 

In the meantime, we need to develop dedicated cosmological simulations
of black hole formation and early growth that can aid the
interpretation of these data.  The suggestion that the accretion rate
of massive black holes depends on their environment, (through the host
halo and its cosmic bias) must be tested with cosmological simulations
that implement physically-motivated accretion and feedback
prescriptions. We also need to derive predictions for the occupation
fraction of black holes in galaxies based on black hole formation
models (Bellovary et al. in prep.).  This will be a significant
improvement over current simulations of black hole cosmic evolution
that typically place black holes in halos growing above some threshold
mass, typically $\sim 10^{10}$ \msun, leading to a trivial occupation
fraction function.  There is no strong physical reason to believe that
all and only galaxies with mass $>10^{10}$\msun~ host massive black
holes in their centers. 

 In this paper we focused on the very high-redshift Universe, $z\simeq 6$.
Although this redshift range is not a special place, the concurrence of
theoretical arguments and observational constraints allow us to 
make simplifying assumptions that are not expected to be valid 
at later times. For instance,  a timescale argument requires 
black holes to grow fast to reach the masses probed by current luminosity functions. 
In turn, this argument, coupled to current observational constraints suggest 
that the most luminous quasars accrete close to Eddington,
and that both active fraction and occupation fractions must be of order unity at the high-luminosity, high-mass 
end. This is not true at $z=2$, where more variables enter into play, and make 
the analysis less constraining.  An example of a detailed study that connects the 
mass function of black holes derived from the Press \& Schechter formalism to that 
derived from the luminosity function is presented in Croton (2009).

\section*{Acknowledgments}
We thank K. G\"ultekin, T. Lauer and D. Richstone for insightful comments on the manuscript. 
MV acknowledges support from SAO Award TM9-0006X and NASA award ATP NNX10AC84G. 
DPS acknowledges support from the STFC through the award of a Postdoctoral 
Research Fellowship.


\bsp

\label{lastpage}

\end{document}